\documentclass{an}

\usepackage{graphicx}
\usepackage{times}
\usepackage{amssymb}

\newcommand\pjet{P_{\rm jet}}
\newcommand\vvec{\mathbf{v}}
\newcommand\bhat{\mathbf{b}}
\newcommand\eg{e.g.~}
\newcommand\etal{et al.}

\begin{document}

\Pagespan{1}{}
\Yearpublication{2012}
\Yearsubmission{2012}

\title{AGN feedback in clusters: shock and sound heating}

\author{P.E.J. Nulsen\inst{1}\fnmsep\thanks{\email{pnulsen@cfa.harvard.edu}}
\and B.R. McNamara\inst{1,2,3}}

\titlerunning{Shock and sound heating in clusters}
\authorrunning{P.E.J. Nulsen}

\institute{Harvard-Smithsonian Center for Astrophysics,
60 Garden Street, Cambridge, MA 02138, USA
\and Department of Physics and Astronomy, University of Waterloo,
Waterloo, ON, Canada 
\and Perimeter Institute for Theoretical Physics, Waterloo, ON, Canada}


\abstract{Observations support the view that feedback, in the form of
  radio outbursts from active nuclei in central galaxies, prevents
  catastrophic cooling of gas and rapid star formation in many groups
  and clusters of galaxies.  Variations in jet power drive a
  succession of weak shocks that can heat regions close to the active
  galactic nuclei (AGN).  On larger scales, shocks fade into sound
  waves.  The Braginskii viscosity determines a well-defined sound
  damping rate in the weakly magnetized intracluster medium (ICM) that
  can provide sufficient heating on larger scales.  It is argued that
  weak shocks and sound dissipation are the main means by which radio
  AGN heat the ICM, in which case, the power spectrum of AGN outbursts
  plays a central role in AGN feedback.}

\maketitle

\section{Introduction}

X-ray observations of cool core clusters and groups have shown that
radio outbursts from central AGN have large impacts on their hot
atmospheres (reviewed in \cite{mn07}).  Radio outbursts originating
near the event horizons of supermassive black holes deposit energy on
spatial scales eight orders of magnitude larger.  In the absence of a
heat source, copious amounts of hot gas would be cooling to low
temperatures and forming stars.  Remarkably, powers of AGN outbursts
are comparable to the powers needed to stop the gas from cooling,
signalling that the mechanical power output of the AGN is governed by
feedback (\cite{brm04}; \cite{df06}; \cite{rmn06}; \cite{csw06};
\cite{ss06}).  Cooled or cooling gas can fuel AGN outbursts, while the
outbursts heat the gas, affecting the fuel supply for subsequent
outbursts.  The high incidence of clusters with central cooling times
less than 1 Gyr (\cite{hmr10}) also argues for radio mode AGN
feedback.  While many processes might heat the ICM and prevent the gas
from cooling, without feedback, it is all but impossible to account
for the many systems with very short cooling times, while essentially
no clusters are undergoing catastrophic cooling (\cite{mn12}).

While the broad outline of the feedback cycle seems simple, very
little of the detail is understood.  In this article we focus on some
processes that may heat the gas on scales comparable to the Bondi
radius and larger.  Apart from the discussion of viscosity, much of
this material has been reviewed more thoroughly by McNamara \& Nulsen
(2012).  We focus on the heating effects of weak shocks in section
\ref{sec:shock} and sound waves in section \ref{sec:sound}.  In
section \ref{sec:discuss} we consider how these processes operate
together.

\section{Heating and outburst history} \label{sec:shock}

\subsection{Adiabatic uplift is ineffective}

To prevent the hot gas from cooling and forming stars, the key
requirement is a heat source to replace heat lost by radiation.  It
should be emphasized that adiabatic uplift is ineffective at
preventing gas from cooling.  Extended filaments of cool, low entropy
gas (\eg \cite{wsm10}; \cite{gnd11}) and heavy elements shed by
evolving stars in the central galaxy (\eg \cite{mws10}; \cite{kmc11})
make a strong case for gas uplift in the wakes of radio lobes.
Lifting gas outward adiabatically, into regions where the atmospheric
pressure is lower, will generally extend its cooling time and so can
help to delay the onset of cooling.  However, for abundances and
temperatures in the relevant range, the effect is modest.  For
example, for gas with solar abundances, starting at 3 keV and reducing
the pressure adiabatically by a factor of $\simeq 88$ would reduce its
temperature to 0.5 keV, but only increase its cooling time by $\simeq
36\%$.  For gas with 0.5 solar abundances, the increase in cooling
time is a little under a factor of 2, still well short of what is
required to prevent the gas cooling in the long term (\cite{mn12}).

Furthermore, unless the uplifted gas mixes with it surroundings,
raising its entropy, it is negatively buoyant and will fall back to
where it came from in about one free-fall time.  That is generally
much shorter than the cooling time, so the effect of lifting the gas
is transient.  Metals shed by cluster central galaxies are more
extended than their stars, showing that they have diffused outward from
where they were shed (\cite{rcb05}).  However, if all the uplifted gas
mixed effectively the mean diffusion rate would be far too high, so
most of the gas must fall back almost to where it originates.  Much of
the energy needed to lift the gas is then converted to kinetic energy
and dissipated in the gas, providing a channel for heating
(\cite{gnd11}).

\subsection{Weak shocks} \label{sec:weak}

The thermal energy of the gas within the volume $V$ is
\begin{equation}
E_{\rm th} = {3/2} \int_V p\, dV,
\end{equation}
where $p$ is its pressure.  If the energy deposited by an AGN into
this volume is comparable to or larger than $E_{\rm th}$, then the
fractional pressure increase in $V$ must be large to accommodate the
extra energy.  That would cause the region affected by the jet to
expand supersonically, driving shocks into its surroundings and
rapidly extending the region affected by the outburst.  Similarly,
shocks will be formed if the jet power exceeds $E_{\rm th}$ divided by
the sound crossing time of $V$.  Thus, the region affected by an
outburst must contain a thermal energy prior to the outburst that is
significantly larger than the outburst energy and also larger than the
jet power times the sound crossing time of the region.  Otherwise,
shocks are generated.  Based on simulations, Morsony \etal~(2010)
have found that, under the influence of cluster ``weather'' due to
continuing infall, motion of substructures, etc., the radius of
influence of the central AGN scales with the power of its jet,
$\pjet$, as $R_{\rm influence} \propto \pjet^{1/3}$.  The argument
here suggests that this cannot be the whole story.

A number of systems, such as MS0735.6+7421 (\cite{mnw05}) and NGC~5813
(\cite{rfg11}), show symmetric, large-scale shocks fronts that do not
appear to be affected significantly by the cluster weather.
Furthermore, in NGC~5813 (\cite{rfg11}) and M87 (\cite{fjc07}), there
is clear evidence of multiple cavities and shocks.  The nested shocks
seen in these systems almost certainly require large and sustained
variations in the power of the jet for their formation, consistent
with other evidence of variations in jet power (e.g{.} \cite{wmn07}).
Ripples in the Perseus ICM (\cite{fst06}) also indicate power
variations (though sound may be driven in other ways, \eg
\cite{ss09}).  Note that the small fluctuations seen at large
distances from the cluster centre would have been considerably greater
when they were launched in the region near NGC~1275.  Power variations
with shorter timescales and/or lower amplitude would be too weak to
see now due to the combined effects of damping (section
\ref{sec:sound}) and the decrease in amplitude with increasing radius.
It is noteworthy that the two best observed systems with AGN
outbursts, M87 and Perseus, show multiple weak shocks.  Given the
ubiquitous evidence of AGN variability on a broad range of timescales,
this is probably the norm.  As argued below, the power spectrum of AGN
outbursts plays a critical role in AGN feedback by controlling the
launching of shocks and sound waves.

Although the heating effect of individual weak shocks is minor, their
cumulative impact need not be.  The changes in thermal and kinetic
energy associated with a weak shock may be substantial, but mostly
move on with the shock.  A small entropy increases, $\Delta S$, that
is cubic in the shock strength (\cite{ll59}), is all that remains.
The heat equivalent of the entropy jump is $\Delta Q = T \, \Delta S =
E \, \Delta \ln K$, where $K = kT/n_{\rm e}^{2/3}$ is the entropy
index and $\Delta \ln K$ is the jump of $\ln K$ in the shock.
Expressed as a fraction of the gas thermal energy, $\Delta Q / E =
\Delta \ln K$.  For the innermost shock in M87, at a radius of $\simeq
0.8$ arcmin (3.7 kpc; \cite{fjc07}), the Mach number is $\simeq 1.38$,
giving an equivalent heat input of only $\Delta Q / E \simeq 0.022$.
There is a second shock at about twice the radius and a third shock
that is several times more energetic at a radius of $\simeq 3$ arcmin.
The shock spacings suggest that shocks of similar strength to the
innermost shock are launched every $\sim 2.5$ Myr, while the
cooling time of the gas at 0.8 arcmin is $\simeq 250$ Myr.  The
$\sim100$ shocks launched during one cooling time would add heat
$\Delta Q_{\rm tot}/E \simeq 100 \times 0.022 = 2.2$, more than
enough to replace the energy radiated.  These numbers are
indicative only, but they show that repeated weak shocks alone can
prevent gas near the centre of M87 from cooling (\cite{njf07}).  A
similar argument has been made for weak shock heating in NGC~5813
(\cite{rfg11}).

Weak shocks can prevent gas cooling at the centres of two of the two
nearest, best observed systems.  The ripples in Perseus may well start
as weak shocks launched from near the centre of NGC~1275, where they
prevent the gas from cooling too.  If the best observed systems are
representative, weak shocks can be the primary channel for heating gas
near the centres of all systems.  Because shock strength declines with
distance from the AGN and $\Delta \ln K$ depends steeply on shock
strength, weak shocks become less effective at larger radii.  However,
as the shock strength decreases, sound dissipation increases in
relative importance, probably taking over as the main heating channel,
as outlined below.  The efficiency of weak shock heating, measured by
the fraction of shock energy converted to heat, is generally low, so
that sound dissipation will often make a greater contribution to the
total heating rate.

\section{Plasma viscosity and sound heating} \label{sec:sound}

\subsection{Plasma viscosity}

Although the magnetic pressure in the ICM is typically only $\sim1\%$
of the gas thermal pressure (\cite{ct02}), particle Larmor radii are
ten or more orders of magnitude smaller than their mean free paths,
keeping the particles tied rigidly to the magnetic field.  As a
consequence, transport processes in the ICM are poorly understood.
This issue is worst for thermal conduction, since the heat flux depends
on the structure of the magnetic field on scales comparable to the
electron mean free path.  The field may well vary on scales smaller
than this, in which case the heat flux is not calculable in terms of
local gas properties.  Despite a great deal of work in this area, the
role of thermal conduction in the ICM remains highly uncertain.

By contrast, viscous stresses are often determined locally.  Fluid
motions readily push the dynamically insignificant field around and,
being frozen in, the field generally varies as the fluid moves.  In a
collisionless plasma, the magnetic moment, $m v_\perp^2 / (2 B)$, is
conserved, where $m$ is the particle mass, the strength of the
magnetic field is $B$ and the component of the particle velocity
perpendicular to the field is $v_\perp$.  Thus changes in the magnetic
field cause corresponding changes in the transverse kinetic energy, $m
v_\perp^2/2$, making the particle velocity distribution anisotropic.
The proton-proton collision time is $\tau_{\rm pp} \simeq 700
(kT)^{3/2} n_{\rm p}^{-1}$ yr, where the temperature, $kT$, is in keV
and the proton density, $n_{\rm p}$, is in $\rm cm^{-3}$.  Collisional
relaxation can generally suppress most of the anisotropy caused by
fluid motions, leaving only a small residual effect.  Anisotropy in
the velocity distribution implies anisotropy in the pressure, which
can be expressed as the difference between the transverse and mean
pressures, $\Delta = p_\perp - p$.  When collisional relaxation is
fast compared to the rate of change in $B$, i.e. $\tau_{\rm ii} |dB
/ dt| \ll B$, where $\tau_{\rm ii}$ is the ion-ion collision time, the
pressure anisotropy is given by (\eg \cite{kbr12})
\begin{equation}
\Delta = \tau_{\rm ii} p_{\rm i} \left(\bhat \bhat : \nabla \vvec 
- {1\over3} \nabla \cdot \vvec\right),
\label{eqn:delta}
\end{equation}
where $p_{\rm i}$ is the ion pressure, $\bhat$ is the direction of the
magnetic field and $\vvec$ is the fluid velocity.  The viscous stress
tensor, which is minus the anisotropic part of the total stress
tensor, is then
\begin{equation}
\mathbf{T} = \Delta (3 \bhat \bhat - \mathbf{1}),
\end{equation}
where $\mathbf{1}$ is the $3\times3$ unit matrix.  This is the
Braginskii (1965) form of the viscous stress tensor for a magnetized
plasma.  Anisotropy generated by motion parallel and perpendicular to
the magnetic field simply reflects work done on or by the components
of the particle velocities through the corresponding particle
pressures, $p_\perp$ or $p_\parallel = 3 p - 2 p_\perp$.  In a uniform
magnetic field parallel to the $z$ axis, with gas motions only
parallel the field, $T_{zz}$ must have the same value as it does in
the absence of a magnetic field.  This requires $\tau_{\rm ii} p_{\rm
  i}$ to equal the viscosity of an unmagnetized plasma, i.e. the
``Spitzer'' viscosity.  Thus, although the viscous stresses depend on
the direction of the magnetic field, they are similar in magnitude to
the stresses in the absence of a magnetic field.

\subsection{Sound heating} \label{sec:damp}

Calculation of sound damping for a magnetized ICM is much the same as
in the field free case.  We assume that the unperturbed magnetic field
is uniform and parallel to the $z$ axis.  For sound with wavevector
$\mathbf{k} = (k_x, k_y, k_z)$, the damping rate of the amplitude is
\begin{equation}
{1\over6} \nu k^2 \left(1 - 3{k_z^2\over k^2}\right)^2,
\label{eqn:damp}
\end{equation}
where $\nu = \tau_{\rm ii} p_{\rm i} / \rho$ is the kinematic
viscosity and $\rho$ is the gas density.  For sound waves travelling
parallel to the magnetic field ($k_z^2 = k^2$), this is identical to
the damping rate with no magnetic field, $2\nu k^2/3$, as we should
expect.  For sound travelling perpendicular to the field ($k_z = 0$),
the damping rate is $\nu k^2 / 6$, a factor of 4 smaller.  Note that
for $k_z^2 = k^2/3$, i.e.~for sound waves inclined at $\simeq
54.7^\circ$ to the magnetic field, the damping rate is zero.  In this
direction sound waves have the same effect on $p_\perp$ and
$p_\parallel$, so they generate no anisotropy, hence no viscous
damping.

Due to cluster weather, the field is likely to be fairly chaotic, with
structure on scales down to 10 kpc or less (\eg \cite{gdm10}).  If
the field is isotropic on average, we can average the damping rate
(\ref{eqn:damp}) over the sphere to obtain the mean value of $2 \nu
k^2 / 15$, one fifth of the damping rate with no magnetic field.
Fabian \etal~(2005) have shown that viscous damping of sound is a
viable mechanism to prevent gas cooling in the Perseus cluster
and Abell 2199.  Their models used a viscous damping rate that is one
half of the mean rate calculated here.  As they discuss, the power
spectrum of the AGN outbursts is critical for the heating rate, since
it controls the spectrum of sound generated by outbursts and the
dissipation rate is sensitive to the frequency.  Corresponding to the
average damping rate, the dissipation length for sound power is
\begin{equation}
120 \left(\omega \over \omega_{\rm s}\right)^{-2}
\left(n_{\rm e} \over 0.03 \rm\ cm^{-3}\right)  
\left(kT \over 5 \rm\ keV\right)^{-1}\rm\ kpc,
\label{eqn:dislen}
\end{equation}
where the angular frequency unit, $\omega_{\rm s} =
2.36\times10^{-14}\rm \ s^{-1}$, gives a wavelength 10 kpc for a gas
temperature of 5 keV (cf.~\cite{frt05}).  This is comparable to the
size of cluster cool cores.  If the thermal conductivity is close to
its field free value, conduction would boost the dissipation rate
significantly.  However, viscous dissipation alone appears sufficient
to heat the ICM in Perseus and Abell~2199.

Schekochihin \etal~(2005) note that, when the anisotropy falls outside
the range $-2p_B/3 < \Delta < p_B/3$, for $p_B = B^2/(8 \pi)$, the
plasma is prone to firehose or mirror instabilities that grow much
faster than viscous damping.  For $p_B \ll p$, as in clusters, this
condition is more restrictive than the one preceding equation
(\ref{eqn:delta}).  It requires $\omega \tau_{\rm ii} A \lesssim 0.5
p_B / p$, where $A$ is the amplitude of the fractional density
fluctuations.  If the anisotropy did get too large, sound damping
could be much faster than the viscous rate and sound heating even more
effective --- probably too effective, although some sound energy would
then be converted to magnetic field rather than being thermalized
(\cite{sck05}).

\section{Discussion} \label{sec:discuss}

Variations in the power of the jet launched from the AGN initiate a
succession of weak shocks that originate from somewhere near the Bondi
radius (disregarding the small amount of gas within the Bondi sphere).
The distribution of gas close to the radio lobes is generally far from
spherical and the shock strength varies over the surface of spheres
(\eg \cite{fnh05}; \cite{mjd12}).  However, shock strength generally
decreases radially.  As the shocks weaken, at some point they can be
regarded as sound waves.  Although the power spectrum of the sound is
far from monochromatic, for the purpose of discussion here, we assume
that there is a representative frequency, $\omega$, where $2 \pi /
\omega$ is typical of the time interval between shocks.

Heating by weak shocks (section \ref{sec:weak}) depends on the
fractional pressure jump, $\delta p / p$, as $(\delta p /p)^3$,
whereas sound damping heats the gas at a rate proportional to the
square of the sound amplitude.  As a result, shock heating is more
significant at small radii where the pressure disturbance is larger,
but as the disturbance decreases with increasing radius, sound
dissipation will overtake it at some point.  If the mean kinematic
viscosity is $\nu/5$ (section \ref{sec:damp}) and the fractional
amplitude of the sound pressure disturbance is related to the pressure
jump by $A_p = 0.5 \delta p / p$, the ratio of the shock heating rate
to the sound heating rate is (\cite{mn07})
\begin{eqnarray}
\lefteqn{{2 s^2 \over \pi \nu \omega} \left(\delta p \over p\right)
\simeq 12.6} \nonumber \\ &&
\left(\omega \over \omega_{\rm s}\right)^{-1}
\left(n_{\rm e} \over 0.03 \rm\ cm^{-3}\right)  
\left(kT \over 5 \rm\ keV\right)^{-3/2} \left(\delta p \over p\right)
\end{eqnarray}
in the units of equation (\ref{eqn:dislen}).  If the entropy or the
effective frequency is high (for wavelengths comparable to the proton
mean free path or smaller), sound damping might rival shock heating,
even for $\delta p/p \sim 1$, but low entropies in cool cores favour
weak shock heating.  Weak shock heating probably dominates close to
the Bondi radius, while sound damping is more effective on larger
scales.  For both processes, the power spectrum of AGN outbursts plays
a central role in determining the heating rate.  Clearly, we need to
understand how that is governed.

\section{Conclusions}

Low rates of gas cooling and star formation in the central galaxies of
many groups and clusters are best explained by feedback from radio
AGN.  This view is supported by observations of the impacts of radio
outbursts in hot atmospheres.  Jet powers vary significantly on a wide
range of timescales, causing a succession of shocks to be launched
from near the radio lobes.  The shocks weaken into sound waves on
larger scales.  For the best observed, nearby systems, weak shocks
provide sufficient heat near the AGN to prevent gas from cooling.

In contrast to thermal conduction, the Braginskii viscosity generally
provides a well-defined, local, viscous stress that only depends on the
direction of the magnetic field in the weakly magnetized ICM.  It
determines a viscous damping rate for sound that is a factor of five
smaller than that for unmagnetized plasma when the field is isotropic
on average.  This level of sound damping is sufficient to prevent gas
from cooling throughout the cool core of the Perseus cluster.  The
combination of weak shock heating on small scales and sound damping on
larger scales plausibly provides the primary means by which AGN energy
heats the ICM and prevents gas from cooling.  If so, the power
spectrum of AGN outbursts plays a central role in AGN feedback and
needs to be better understood.

\section*{Acknowledgement}

This work was supported by NASA grant NAS8-03060.

\newcommand\nat{Nature~}

\end{document}